# (1173) Anchises – Thermophysical and Dynamical Studies of a Dynamically Unstable Jovian Trojan

Horner, J.[1], Müller, T. G.[2] & Lykawka, P. S.[3]

[1] Department of Astrophysics and Optics, School of Physics, University of New South Wales, Sydney 2052, Australia.
[2] Max-Planck-Institut für extraterrestrische Physik (MPE), Giessenbachstrasse, 85748, Garching, Germany
[3] Astronomy Group, Faculty of Social and Natural Sciences, Kinki University, Osaka, Japan.
Contact: j.a.horner@unsw.edu.au

**Abstract**
We have performed detailed thermophysical and dynamical modelling of the Jovian Trojan (1173) Anchises. Our results show that this is a most unusual object. By examining observational data taken of Anchises by IRAS, Akari and WISE at wavelengths between 11.5 and 60 microns, together with the variations in its optical lightcurve, we find that Anchises is most likely an elongated body, with an axes-ratio, a/b, of around 1.4. This results in calculated best-fit dimensions for Anchises of 170x121x121 km (or an equivalent diameter of 136 +18/-11 km). We find that the observations of Anchises are best fit by the object having a retrograde sense of rotation, and an unusually high thermal inertia in the range 25 to 100 $Jm^{-2}s^{-0.5}K^{-1}$ (3-σ confidence level). The geometric albedo of Anchises is found to be 0.027 (+0.006/-0.007). Anchises therefore has one of the highest published thermal inertias of any object larger than 100 km in diameter, at such large heliocentric distances, as well as being one of the lowest albedo objects ever observed. More observations (visual and thermal) are needed to see whether there is a link between the very shallow phase curve, with almost no opposition effect, and the derived thermal properties for this large Trojan asteroid. Our dynamical investigation of Anchises' orbit has revealed it to be dynamically unstable on timescales of hundreds of millions of years, similar to the unstable Neptunian Trojans 2001 $QR_{322}$ and 2008 $LC_{18}$. Unlike those objects, however, we find that the dynamical stability of Anchises is not a function of its initial orbital elements, the result of the exceptional precision with which its orbit is known. Our results are the first time that a Jovian Trojan has been shown to be dynamically unstable, and add further weight to the idea that the planetary Trojans likely represent a significant ongoing contribution to the dynamically unstable Centaur population, the parents of the short-period comets. The observed instability (fully half of all clones of Anchises escape the Solar system within 350 Myr) does not rule out a primordial origin for Anchises, but when taken in concert with the result of our thermophysical analysis, suggest that it would be a fascinating target for future study.

**Keywords:**
minor planet, asteroids: general

**Introduction**
Since the discovery of the first Jovian Trojan, (588) Achilles, back in 1906, the origin and nature of planetary Trojans has been widely debated (e.g. Gomes, 1998; Fleming & Hamilton, 2000; Nesvorný & Dones, 2002; Kortenkamp, Malhotra & Michtchenenko, 2004; Morbidelli et al., 2005; Lykawka et al., 2009; Lykawka et al., 2010). To date, almost 5,000 Jovian Trojans have been discovered, spanning a wide range of orbital eccentricities and inclinations. Other planets, too, have been found to host Trojans – Neptune is accompanied by eight Trojan companions, whilst Mars has four[1]. Recently, the first Trojan companion to the Earth was discovered, though that object seems likely to be a recently captured, rather than long term member of the Solar system's Trojan population (Connors, Weigert & Ceillet, 2011).

---

[1] Numbers taken from http://www.minorplanetcenter.net/iau/lists/Trojans.html on 11th August 2011.

The nature of the planetary Trojans may well be the key to unravelling details of the formation and evolution of our planetary system (e.g. Morbidelli et al., 2005; Lykawka et al., 2009, 2011). The currently accepted paradigm is that the planetary Trojans are a captured, rather than truly primordial, population (e.g. Lykawka et al., 2009). Rather than forming in situ, it is thought that the Jovian and Neptunian Trojan populations were captured as a result of the migration of those giant planets toward the latter stages of their formation. Although some models propose that that capture was the result of a chaotic and unstable migration of the giant planets, featuring their making destabilising mutual mean-motion resonance crossings, and potentially resulting in a cascade of rocky and icy material towards the terrestrial planets (the putative Late Heavy Bombardment) (Morbidelli et al., 2005), other work has shown that the populations could just as easily be produced as a result of smooth, gentle migration of the giant planets (Lykawka et al., 2009, Lykawka & Horner, 2010).

Regardless of the precise details of the Trojan capture, it is clear that study of these interesting relics of planetary formation can lead to significant advances in our understanding of the details of planetary formation and migration. The capture model of Trojan origin is currently the only way to explain the wide range of orbital eccentricities and inclinations displayed by the objects, but it also makes another, implicit prediction. Were the Trojans captured in this way (and irrespective of whether that capture was the result of chaotic or smooth planetary migration), then objects would have been captured to the Trojan clouds on orbits covering not just a wide range of orbital eccentricities and inclinations, but also a wide variety of dynamical stabilities, from the tightly bound objects stable on timescales far greater than the age of our Solar system to loosely held members that would escape the clouds on far shorter timescales. Between these two extremes, there would clearly be a range of objects captured onto orbits of intermediate stability, leading to the gradual ongoing decay of the Trojan populations, and, in turn, an ongoing flux of Trojans into the Centaur and short-period comet populations (e.g. Horner & Lykawka, 2010a, c). But the question is – do such unstable Trojans exist?

For the Neptune Trojans, we have recently shown that both 2001 QR$_{322}$ and 2008 LC$_{18}$ (two of the eight Trojans known) may well be dynamically unstable on timescales of hundreds of millions of years (Horner & Lykawka, 2010b; Horner et al., 2012) – a result that is entirely compatible with them having been captured during the epoch of planetary migration, and remained as Trojans ever since. But what of the Jovian Trojans? Are any of those objects similarly dynamically unstable?

Historically, a number of studies have examined the stability of the Jovian Trojans. Several analytical, semi-analytical and early numerical studies investigated the various regions of stability, the influence of initial orbital parameters (e.g., eccentricities and inclinations) and the effects of secular resonances on Trojan asteroids (e.g., Freistetter 2006 and references therein). In addition, numerical simulations and dynamical mapping have been used to explore the fine structure of Trojan motion, including orbital integrations of fictitious and real objects over the age of the Solar system (Levison, Shoemaker & Shoemaker 1997; Robutel & Gabern 2006; Melita et al. 2008; Stacey & Connors 2008; Lykawka & Horner 2010). In particular, Robutel & Gabern (2006) investigated the dynamics of Anchises in some detail, finding that the object lies close to a number of secular resonances, which potentially suggests it is evolving on a "stable chaotic" orbit. When the current best-fit orbital solution for Anchises is considered in the context of models of the origin and dynamical evolution of Jovian Trojans (e.g. Morbidelli et al. 2005; Lykawka & Horner 2010), it is worth noting that its orbit seems to lie close to, but outside, the region of long-term stability determined by those models.

(1173) Anchises was the 9$^{th}$ Jovian Trojan to be discovered, and was first recorded in 1930, over 80 years ago. It librates around the Jovian L5 Lagrange point, trailing Jupiter in its orbit around the Sun, and is categorised as a P-type asteroid (following Tholen, 1984), a result supported by the B – V and V – R colours measured by Fornasier et al. (2007). Its size and albedo have been determined from thermal infrared observations via radiometric techniques (Cruikshank, 1977; Morrison, 1977; Morrison & Zellner, 1979; Tedesco et al., 2002; Usui et al., 2011). The derived equivalent diameters range between 64 and 159 km, whilst the geometric visual albedo has been estimated to lie between 0.02 and 0.05. One of the reasons for the large spread in calculated diameter values might be related to the results being based on single-epoch observations of Anchises, which has been observed to display a large amplitude light curve. French (1987) found that Anchises rotates with a period of 11.84 hours, and that its light curve displays a peak-to-peak variation of 0.57 mag, indicating that it is a very elongated body. The most recent study, by Usui et al. (2011), included multiple-epoch *Akari* observations carried out at 18 microns. The derived equivalent diameter presented in that work is 120.5 ± 2.9 km, and the obtained geometric visual albedo (assuming that the object has an absolute magnitude, $H$, of 8.89, a value we use throughout this work; Lagerkvist et al. (2001)) is 0.035 ± 0.002. The observations detailed by French (1987) also suggest that Anchises displays an unusual surface texture, and is probably far less rough than the great majority of asteroids (based on its lacking any noticeable opposition effect and the small phase coefficient). Fornasier et al. (2007) found that Anchises has the lowest spectral slope among all L5 Trojans investigated to date, and confirmed the low value of 3.8% / $10^3$Å, given in Jewitt & Luu (1990). However, a more sophisticated thermophysical model study is required in order to validate the suspicion that Anchises has a very smooth surface, potentially even bare rock, with high thermal inertia. If Anchises truly is an unusually smooth body, then it could well be the case that its true diameter could be significantly greater than that stated above. It is clearly timely, therefore, to revisit our understanding of Anchises, in light of recent observations.

In this work, we present the results of detailed thermophysical and dynamical studies of Anchises. We first discuss the physical properties of Anchises, in section two. In section three, we describe the simulations we performed to investigate the dynamical behaviour of Anchises, before presenting and discussing our results in section four. Finally, we draw our conclusions, and discuss possible future work, in section five.

**Thermophysical analysis of (1173) Anchises**

(1173) Anchises has been observed multiple times at thermal infrared wavelengths. We have collected the available data, and translated them into monochromatic flux densities at the given reference wavelength (see Table 1). The IRAS data (3 visits, each with a 4-band detection) have been taken from the electronic tables connected to the publication by Tedesco et al. 2002, and have been colour corrected using a model spectral energy distribution that uses the corresponding heliocentric distance of 5.7 AU and a geometric albedo of 4%. The colour-correction terms were 0.83, 0.98, 1.12, and 1.06 at 12, 25, 60 and 100 microns. This correction converts the measured broadband flux into a monochromatic flux density, which can then be used for thermophysical modelling. All IRAS 12 and 100 micron measurements have been skipped, either due to their unknown measurement uncertainties, or due to their having SNR below 5 (see also the analysis by Tedesco et al. 2002 presented in the electronically available tables SIMPS.FP208A.dat and SIMPS.FP208B.dat).

The Akari data (5 detections at 18 microns, none at 9 micron) are taken (after appropriate colour-correction) from the list of measured fluxes that were used as input for the AcuA-catalogue (Usui et al. 2011). The WISE data are described by Mainzer et al. 2011, and are available from the WISE archive (http://irsa.ipac.caltech.edu/applications/wise/). We calibrated the WISE datasets using the provided Vega model spectrum, and applied appropriate colour-corrections (W2: 1.087; W3: 1.115; W4: 0.982) for the object's 4.93 AU heliocentric distance at the time of the observations. The WISE-W1 and W2 data have been skipped due to significant contributions from reflected light (we calculated that even the signal in the W2-band at 4.6 micron includes up to 40% reflected light, depending on surface roughness and thermal inertia). The derived monochromatic flux densities for the W3 and W4 band observations have an associated uncertainty of ±10% (http://wise2.ipac.caltech.edu/docs/release/prelim/expsup/).

| Julian date | λ [μm] | $FD_{cc}$ [Jy] | Error [Jy] | $r_{helio}$ [AU] | Δ [AU] | α [°] | L/T | telescope/ instrument |
|---|---|---|---|---|---|---|---|---|
| 2445510.65537 | 25.00 | 1.83 | 0.42 | 5.67095 | 5.57780 | 10.32 | T | IRAS |
| 2445510.65537 | 60.00 | 0.88 | 0.19 | 5.67095 | 5.57780 | 10.32 | T | IRAS |
| 2445518.09699 | 25.00 | 0.91 | 0.19 | 5.66493 | 5.68734 | 10.27 | T | IRAS |
| 2445518.09699 | 60.00 | 0.80 | 0.15 | 5.66493 | 5.68734 | 10.27 | T | IRAS |
| 2445518.02536 | 25.00 | 1.15 | 0.26 | 5.66499 | 5.68629 | 10.27 | T | IRAS |
| 2445518.02536 | 60.00 | 0.83 | 0.16 | 5.66499 | 5.68629 | 10.27 | T | IRAS |
| 2453873.28118 | 18.00 | 0.3776 | 0.0373 | 6.03137 | 5.95544 | 9.65 | T | Akari-L |
| 2453873.34995 | 18.00 | 0.5002 | 0.0426 | 6.03136 | 5.95652 | 9.65 | T | Akari-L |
| 2454264.18722 | 18.00 | 0.5826 | 0.0478 | 5.86619 | 5.78817 | 9.97 | T | Akari-L |
| 2454090.00829 | 18.00 | 0.5756 | 0.0476 | 5.95941 | 5.88773 | 9.50 | L | Akari-L |
| 2454090.07727 | 18.00 | 0.7048 | 0.0538 | 5.95938 | 5.88657 | 9.50 | L | Akari-L |
| 2455276.56969 | 11.56 | 0.1812 | 0.0181 | 4.92995 | 4.83373 | 11.66 | L | WISE |
| 2455276.56969 | 22.09 | 0.9915 | 0.0992 | 4.92995 | 4.83373 | 11.66 | L | WISE |
| 2455276.70199 | 11.56 | 0.2651 | 0.0265 | 4.92982 | 4.83154 | 11.66 | L | WISE |
| 2455276.70199 | 22.09 | 1.3928 | 0.1393 | 4.92982 | 4.83154 | 11.66 | L | WISE |
| 2455276.83442 | 11.56 | 0.1591 | 0.0159 | 4.92969 | 4.82934 | 11.66 | L | WISE |
| 2455276.83442 | 22.09 | 1.0952 | 0.1095 | 4.92969 | 4.82934 | 11.66 | L | WISE |
| 2455276.96673 | 11.56 | 0.2494 | 0.0249 | 4.92957 | 4.82715 | 11.66 | L | WISE |
| 2455276.96673 | 22.09 | 1.4951 | 0.1495 | 4.92957 | 4.82715 | 11.66 | L | WISE |
| 2455277.03282 | 11.56 | 0.2152 | 0.0215 | 4.92950 | 4.82605 | 11.66 | L | WISE |
| 2455277.03282 | 22.09 | 1.0382 | 0.1038 | 4.92950 | 4.82605 | 11.66 | L | WISE |
| 2455277.09903 | 11.56 | 0.2641 | 0.0264 | 4.92944 | 4.82496 | 11.66 | L | WISE |
| 2455277.09903 | 22.09 | 1.2820 | 0.1282 | 4.92944 | 4.82496 | 11.66 | L | WISE |
| 2455277.16512 | 11.56 | 0.2571 | 0.0257 | 4.92938 | 4.82386 | 11.66 | L | WISE |
| 2455277.16512 | 22.09 | 1.4705 | 0.1471 | 4.92938 | 4.82386 | 11.66 | L | WISE |
| 2455277.23134 | 11.56 | 0.2378 | 0.0238 | 4.92931 | 4.82276 | 11.66 | L | WISE |
| 2455277.23134 | 22.09 | 1.3838 | 0.1384 | 4.92931 | 4.82276 | 11.66 | L | WISE |
| 2455277.29755 | 11.56 | 0.1702 | 0.0170 | 4.92925 | 4.82166 | 11.66 | L | WISE |
| 2455277.29755 | 22.09 | 0.9842 | 0.0984 | 4.92925 | 4.82166 | 11.66 | L | WISE |
| 2455277.42985 | 11.56 | 0.2727 | 0.0273 | 4.92912 | 4.81947 | 11.66 | L | WISE |
| 2455277.42985 | 22.09 | 1.4965 | 0.1497 | 4.92912 | 4.81947 | 11.66 | L | WISE |
| 2455277.56216 | 11.56 | 0.1996 | 0.0200 | 4.92900 | 4.81728 | 11.66 | L | WISE |
| 2455277.56216 | 22.09 | 1.0722 | 0.1072 | 4.92900 | 4.81728 | 11.66 | L | WISE |
| 2455277.69446 | 11.56 | 0.2700 | 0.0270 | 4.92887 | 4.81508 | 11.66 | L | WISE |
| 2455277.69446 | 22.09 | 1.4869 | 0.1487 | 4.92887 | 4.81508 | 11.66 | L | WISE |

Table 1: Summary of all thermal observations used in this work. The time at which the observation was made (in Julian Date), the reference wavelength (in μm), and the colour-corrected, monochromatic flux densities (with errors) are given together with the heliocentric distance, the observatory-centric distance (in AU), and the phase angle. Here, 'L' means leading the Sun (observations which therefore feature the cold terminator, as seen from Earth, for a retrograde sense of rotation), while 'T' means trailing the Sun. The IRAS data were taken on June 25, and July 2, 1983; the Akari data on May 17, Dec. 20, 2006 and June 12, 2007; and the WISE data on March 21/22, 2010.

Before interpreting the available thermal data, we first revisit the results of French (1987). In that work, she derived a rotation period for (1173) Anchises of 11.6095 ± 0.0036 h, based on a large sample of lightcurve measurements. Her results also revealed that Anchises displays a large lightcurve amplitude of 0.57 ± 0.01 mag. Such a large amplitude can best be explained by Anchises being an elongated body with an axes-ratio, a/b, of about 1.4 (determined as the minimum axes-ratio for a rotating ellipsoidal shape model). Some solutions that feature a more elongated body (i.e. a/b>1.4), featuring specific spin-vector orientations, might also explain the observed lightcurve amplitude and therefore cannot be entirely ruled out. The only viewing geometry that can be rejected with high confidence is the pole-on geometry, which fails to reproduce the large light curve amplitude observed. We used the H-G values of $H_V$= 8.89 mag and G=0.03, taken from Lagerkvist et al. (2001), which are based on all available previously published photometric points and lightcurves. Unfortunately, the uncertainty in the rotation period is too large, and the time interval between observations too long, to verify the single-epoch H-magnitude presented in Fornasier et al. (2007). For the same reasons, it was also impossible to transport the French (1987) lightcurve phases all the way to the WISE epochs. Fortunately, the WISE data do have excellent coverage of the object's full rotation and allow for a direct comparison between the WISE thermal measurements and the French (1987) optical lightcurves.

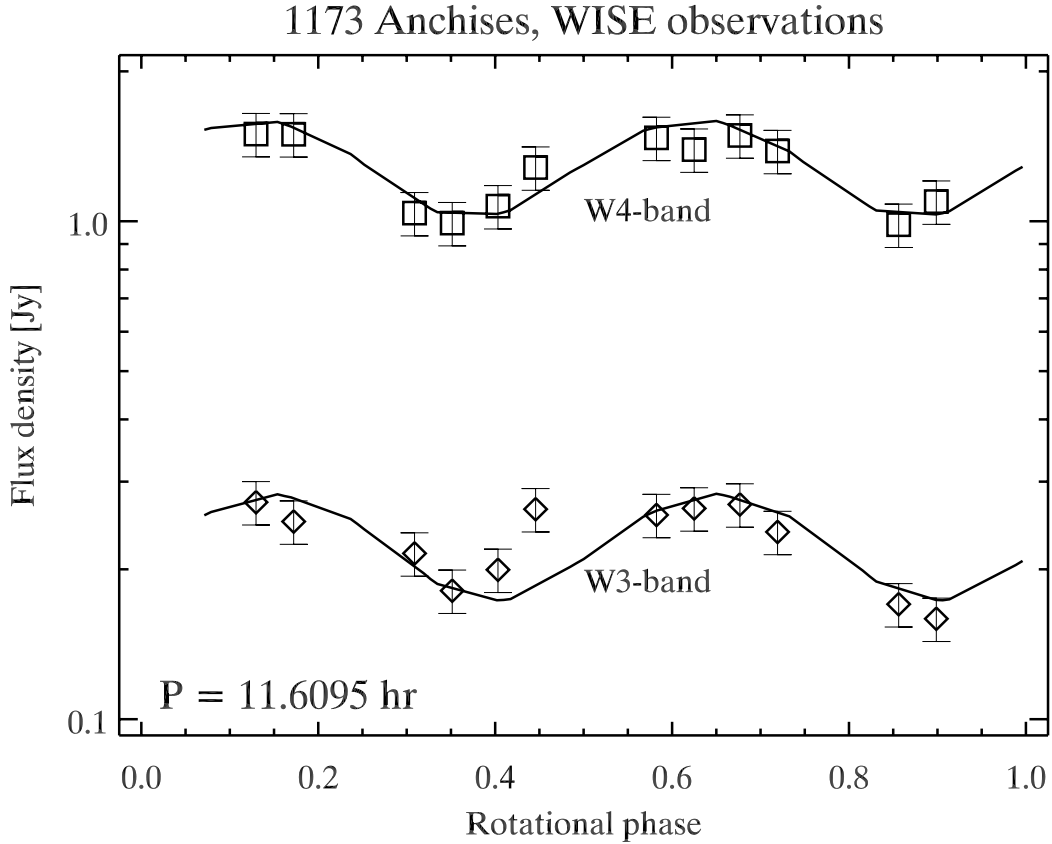

**Figure 1**: The flux densities derived from the two long-wavelength bands of WISE observations[2], as a function of the rotational phase of (1173) Anchises, together with the given ±10% error bars. The observed thermal lightcurve nicely follows the optical lightcurve presented in Figures 1 and 4 of French (1987), indicating that the flux variations are very likely dominated by shape effects and not due to albedo variations on the surface. The absolute flux predictions of our best model solution are shown as solid lines (see explanations in the text). The reason for the outlier in the W3-band, at rotational phase 0.45, is unclear, but that datum is most likely an interloper, partially contaminated by a background star in the crowded field.

The analysis of the available thermal data was carried out using the thermophysical model (TPM) of Lagerros (1996, 1997, 1998), and follows the methodology used to study the approach of the near-Earth asteroids (162173) (1999 JU3; Müller et al. 2011) and (25143) Itokawa (Müller et al. 2005). As a shape model, we used the ellipsoidal shape model of minimal elongation which best matched the optical lightcurve (a/b=1.4, b/c=1.0, $P_{sid}$ = 11.6095 h). The TPM also includes the modelling of Anchises' surface roughness, described by the fraction of the surface covered by craters (f) and the r.m.s. of the surface slopes (ϱ). Our default starting values were f=0.6 and ϱ=0.7, but we also varied ϱ from 0.0 to 1.0 for a scenario in which Anchises' surface was totally covered by craters (i.e. f=1.0). The thermal data are very sensitive to the orientation of the spin-vector. We therefore repeated the analysis for various spin-orientations with $β_{ecliptic}$ (spin-vector) = +/-45°, +/-60°, +/-90°, and tested several different values for the $λ_{ecliptic}$ (spin-vector), all of which are wholly compatible with the observed lightcurves obtained by French (1987). The best match to the WISE data was obtained for a moderately high thermal inertia of around 45 $Jm^{-2}s^{-0.5}K^{-1}$, an equivalent size of 136 +18/-11 km, and a geometric albedo of 0.027 +0.006/- 0.007[3]. A retrograde sense of rotation produced a better match to the WISE data as well as to all available thermal observations. The lowest $\chi^2$-values were obtained for intermediate levels of roughness (0.3 < $\varrho$ < 0.8, f=1.0).

---

[2] Only the two longest wavelength bands are used, since at shorter wavelengths, the light received from (1173) Anchises still contains a significant component of reflected light. At the longer wavelengths, however, the level of reflected light is negligible, and as a result, the observations are essentially solely of thermal radiation.

It is worth noting, here, that approximately half of our observations were taken before opposition (i.e. leading the Sun), with the remainder taken after opposition (trailing the Sun). This, in turn, means that the thermal fluxes include contributions from the cold terminator (in one case) and from the warm terminator (for the other case). The thermophysical model analysis and associated $\chi^2$ analysis is therefore sensitive not only to the thermal inertia, but also to the sense of motion (Müller, 2002). The lower $\chi^2$-values (Fig.2) for a retrograde rotation mean that the measurements after opposition (trailing the Sun) feature the imprint of a warm terminator, whilst the rest of the data lack that signature. Datasets that are limited in wavelength and/or phase-angle coverage usually suffer from the degeneracy between the observable effects of thermal inertia (decreasing day-side temperatures) and roughness (increasing temperatures). Overall, the broad coverage of observations before and after opposition, the wide wavelength-coverage (from 11 to 60 microns), and the excellent coverage in rotational phase acted to significantly reduce this degeneracy effect, and we were therefore able to solve for the size, albedo, thermal inertia and sense of rotation of Anchises, in a manner very similar to the work of Müller et al. (2011).

The $\chi^2$ analysis (Fig. 2) reveals that the ellipsoidal shape model fits the WISE data extremely well (with a reduced $\chi^2$ well below 1; see solid lines in Fig. 1 and the WISE-related solid line in Fig. 2. However, the analysis of WISE data alone would still allow a pro- or retrograde sense of rotation. By contrast, when all the data is combined, the retrograde sense of motion is clearly favoured (solid line related to "all data" in Fig. 2). Such combination does, however, require that Anchises possess a yet higher thermal inertia, of approximately 60 $Jm^{-2}s^{-0.5}K^{-1}$. On the other hand, the $\chi^2$–values associated with this combined data are somewhat larger, which is the direct result of the poorer fit to the Akari and IRAS data. Despite this, we note that the very high $\chi^2$–values that result from a prograde solution for the combined data set means that such a solution can be excluded with high probability. We believe that the raised $\chi^2$–values in the combined data are simply the result of the uncertainties in the rotation period, which mean that it is only possible to obtain a good match at a single epoch, with the rotating ellipsoidal model falling out of phase for the other epochs considered. Nevertheless, using all of the available data (combined with a rotating ellipsoid or a rotating sphere) yields very similar values for the thermal inertia, size, and albedo, and also the preference for the retrograde sense of rotation.

---

[3] The error on the diameter of Anchises has been estimated through the quadratic combination of the absolute flux uncertainty inherent in the WISE data (±10% in flux, corresponding to a ±5% error in the objects size) with the error resulting from the 3-σ confidence range in thermal inertia (from 25 to about 100 $Jm^{-2}s^{-0.5}K^{-1}$), the shape of the asteroid, and the uncertainty in the spin-vector orientation (our analysis was repeated for values of $\beta_{ecliptic}$ ranging from -45° to -90° for each of several different $\lambda_{ecliptic}$ for the spin-vector). The uncertainty in Anchises' geometric albedo is directly dependent on the uncertainty in the value of H used in its calculation. We based our error calculations on an admittedly slightly arbitrary estimate that the value of H is only accurate to +/-0.2 magnitudes. If we use $H_V$=9.25 mag, the value given by French (1987), then the geometric albedo obtained would drop to an extremely low value of 0.018, slightly outside our error boundaries.

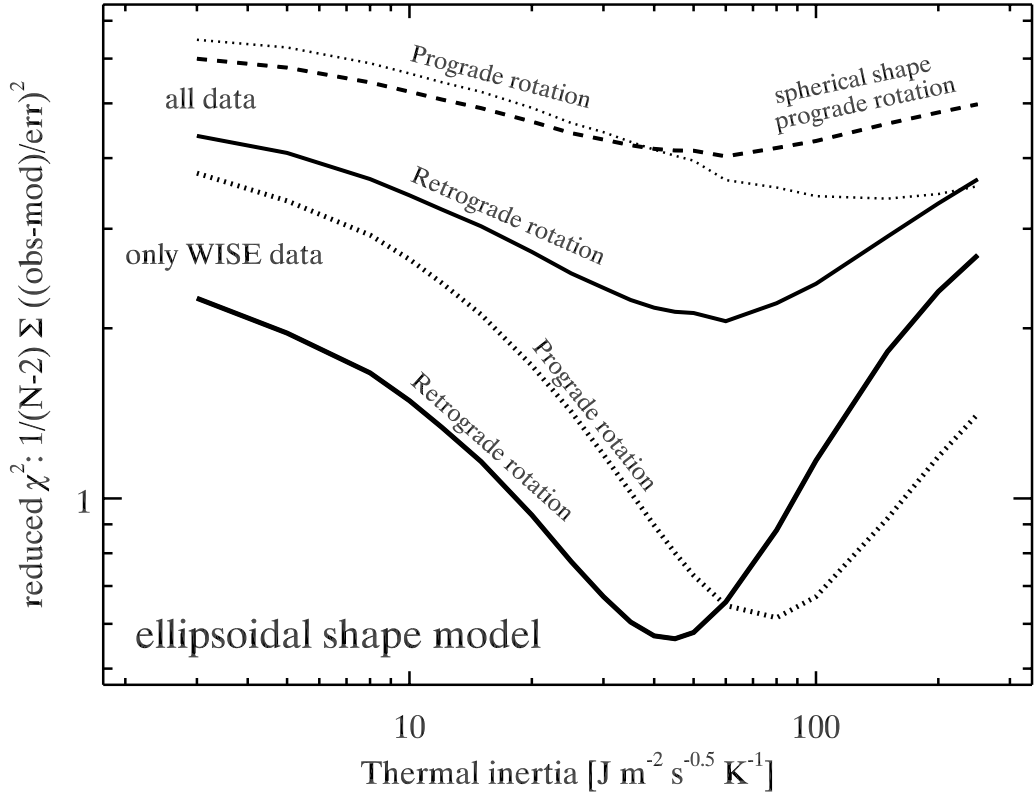

**Figure 2:** $\chi^2$-analysis of the fit between the observed fluxes (Table 1) and the predictions of our thermophysical model for WISE data alone (lower curves) and for all data combined (upper curves). If a simple spherical shape is considered instead of an ellipsoidal shape, the model produces a minimum at similar thermal inertias, but with significantly increased $\chi^2$ values (dashed line).

The derived equivalent diameter corresponds to an ellipsoidal shaped body of dimension 170x121x121 km. These values compare reasonably well with previous results (Table 3, below), especially when one considers that previous studies used only spherical models, and that the object's cross-section is changing significantly with rotation. However, there is additional reason for discrepancy. The relatively large thermal inertia we have measured for Anchises was not considered in the simple thermal models that were used before. The derived thermal inertia is sufficient to transport a significant amount of heat to the night side, and simple models therefore underestimate the size of the object. Our derived albedo agrees very well with previous works, but this is linked to *H*, the absolute magnitude of the object, the accepted value of which has not changed in recent years, although this might need to be revisited in light of changing aspect angles.

| D (km) | $p_V$ | Observations & Notes | Authors |
|---|---|---|---|
| 80±20 | 0.02 | Ground-based Q-band | Cruikshank, 1977 |
| 64 | - | TRIAD file (based on best guess albedo) | Bowell et al., 1979 |
| 92 | 0.047 | Standard radiometry | Morrison & Zellner, 1979 |
| 135±16 | 0.026±0.006 | IRAS asteroid and comet survey | Tedesco, 1986 |
| 126.27±10.7 | 0.0308±0.006 | IRAS, 3 observations | Tedesco et al., 2002 |
| 109-159 | 0.019-0.041 | IRAS, single-band values | (SIMPS.FP208A.dat) |
| 141.2±1.6 | 0.028±0.001 | IRAS | Fernández et al. 2003 |
| 120.49±2.91 | 0.035±0.002 | Akari, 5 observations at L-band | Usui et al., 2011 |
| 114.4±8.0 | 0.038±0.009 | WISE, 7 observations | Grav et al., 2011 |
| 170x121x121 | 0.027 | IRAS, Akari, WISE | This work |

**Table 2:** Previous results, in chronological order, estimating the diameter, $D$, and geometrical V-band albedo ($p_V$) of Anchises. We note that our models involving an equivalent size of 136 +18/-11 km and a geometric albedo of 0.027 +0.006/-0.007 provided the best fit to the WISE data, and are wholly compatible with the range of values detailed in these earlier works.

French (1987) mentioned that Anchises displayed unusual values for the linear phase coefficients for such a dark object, reflected in a moderate opposition effect, and speculated that it may have an unusually smooth surface. Our moderately high thermal inertia (at these very low temperatures, beyond 5 AU from the Sun) indicates that the surface cannot be covered by a thick, low conductivity dust regolith, as is the case for most large main-belt asteroids (Müller & Lagerros 1998, 2002; Müller et al. 1999). It might well be that small values for the linear phase coefficient (absence of strong opposition effects) are, in general, an indication of higher thermal inertias and the lack of a thick, low-conductivity dust regolith. This result could well be applicable to other P-type asteroids (Shevchenko et al. 1997). Grav et al. (2011) listed a beaming parameter, $\eta$, of 0.88±0.12 for Anchises, which led us to consider whether such an $\eta$-value might be typically indicative of an unusually high thermal inertia. It this context, it is interesting to note that the Trojan asteroid (617) Patroclus has previously been found to feature a similar beaming parameter to that obtained for Anchises in this work, with $\eta$ =0.90+/-0.08. Furthermore, Mueller et al. (2010) calculated a thermal inertia for (617) Patroclus of 20 +/- 15 $Jm^{-2}s^{-0.5}K^{-1}$. This value agrees with our 3-sigma confidence interval for Anchises' thermal inertia. French (1987) speculated whether their observations suggested that Anchises' has a smoother (and therefore possibly much younger) surface than is the norm. Indeed, the $\chi^2$-solutions for smoother surfaces (featuring rms-surface slopes < 0.3) point to lower thermal inertias of around 20-30 $Jm^{-2}s^{-0.5}K^{-1}$, but the corresponding $\chi^2$-minima are significantly higher. Despite the uncertainties in the shape and spin-vector orientation of Anchises, the $\chi^2$-picture originating from our dataset points to "normal roughness level" (rms-slopes in the range 0.3 to 0.8 for a 100% cratered surface), but a relatively high thermal inertia (at the given large heliocentric distance). Anchises was found to show a very shallow phase curve and almost no opposition effect (French 1987), but more observations (both visual and thermal), in combination with a better shape and spin-vector solution, are needed in order to establish a clear link between the peculiar reflected light effects observed for Anchises and its inherent roughness and thermal inertia properties.

**Simulations**
In order to study the long-term dynamical behaviour of (1173) Anchises, we performed detailed simulations using the *Hybrid* integrator within the *n*-body dynamics package *MERCURY* (Chambers, 1999). Following our earlier studies of known Solar system bodies (e.g. Horner et al., 2004a, b; Horner & Lykawka 2010b, Horner et al., 2012), we evenly distributed massless "clones" of Anchises evenly across the 3-σ error ellipse in the objects semi-major axis, eccentricity and inclination. In this manner, we created a grid of *27x27x27* test particles in *a-e-i* space, centred on the nominal orbit for the object at epoch 2455600 (as detailed in Table 3), as obtained from the AstDyS website[4]. Since the object has been regularly observed since its discovery in 1930, its orbit is significantly more tightly constrained than those of other objects we have studied (2001 QR$_{322}$ - Horner & Lykawka, 2010b; 2008 LC$_{18}$ – Horner et al., 2012), resulting in a suite of clones spread across a very narrow region of element space. These mass-less test particles were then followed, under the gravitational influence of the Earth, Mars, Jupiter, Saturn, Uranus and Neptune (moving on the orbits detailed in the Jet Propulsion Laboratory's DE405 ephemeris), for a period of 4 Gyr, with an integration timestep, outside of close encounters between test particles and the massive bodies, of 36.525 days. Over time, the "cloud" of clones dispersed, and any test particle that reached a heliocentric distance of 1000 AU was considered to have been ejected from the Solar system, and was removed from the on-going integration. Similarly, any test particle that collided with one of the massive bodies in the integration (the planets, or the central body, the Sun) was removed from the integration. The times at which these ejection or collision events occurred was recorded, allowing the decay of the test population to be recorded as a function of time.

---
[4] http://hamilton.dm.unipi.it/astdys

|   | Value | 1σ variation | Units |
|---|---|---|---|
| a | 5.30962 | 1.765x10$^{-7}$ | AU |
| e | 0.138684 | 1.115x10$^{-7}$ |  |
| i | 6.913 | 1.1x10$^{-5}$ | deg |
| Ω | 283.899 | 9.926x10$^{-5}$ | deg |
| ω | 40.854 | 1.084x10$^{-4}$ | deg |
| M | 333.927 | 5.098x10$^{-5}$ | deg |

**Table 3:** The best-fit orbital elements for (1173) Anchises, together with their associated 1σ errors, at epoch JD 2455600, as obtained from the AstDyS website. These values were used to create a suite of 27x27x27 massless test particles, distributed in *a-e-i* space, centred on the nominal orbit, as detailed in the main text.

**Results and Discussion**

By the end of our simulations, at the 4 Gyr mark, all but 224 of the test particles had been removed from the Solar system, either through a collision with one of the planets, the central body, or reaching a heliocentric distance of 1000 AU. Of the 19459 test particles that were removed in this way, 409 collided with Jupiter, 53 collided with Saturn, 62 collided with the Sun, and 2 hit Uranus. No objects were recorded impacting the Earth, Mars, or Neptune. The remaining 18933 test particles were ejected from the system by reaching 1000 AU. At the end of the simulation, the remaining 224 test particles were all still moving on orbits within the Jovian Trojan cloud.

Why, then, does (1173) Anchises appear to be highly unstable? As was noted in Robutel & Gabern (2006), the orbital motion of (1173) Anchises is strongly influenced by the overlap of two particular secular resonances (plus the 1:1 mean-motion resonance that describes the Trojan motion itself). Those resonances are described by $s = s_6$ and $-5s + 3s_6 + 4g_5 - 2g_6 = 0$, where the resonant quantities refer to the nodal precession frequency of the object ($s$) and that of Saturn ($s_6$), and the perihelion precession frequency of Jupiter ($g_5$) and Saturn ($g_6$). The fact the majority of the population of clones of (1173) Anchises was lost is not, however, particularly unsurprising, since the mean dynamical lifetime for objects in the Centaur region (moving on unstable orbits between the orbits of Jupiter and Neptune – see e.g. Horner et al., 2003) is typically somewhat less than 3 Myr (e.g. Horner et al., 2004a, b), which is already a factor of 1000 times shorter than the integration timescale. However, as can be seen in Horner et al., 2004a, objects with perihelia near the orbit of Jupiter are typically removed from the Solar system on timescales an order of magnitude shorter still. Once a Jovian Trojan leaves the Trojan cloud, it enters the realm of the Jupiter-family comets and the Centaur population (the boundary between which is still poorly defined, e.g. Horner et al., 2003; Gladman, Marsden & Vanlaerhoven, 2008), and can be expected to be removed from the system remarkably quickly.

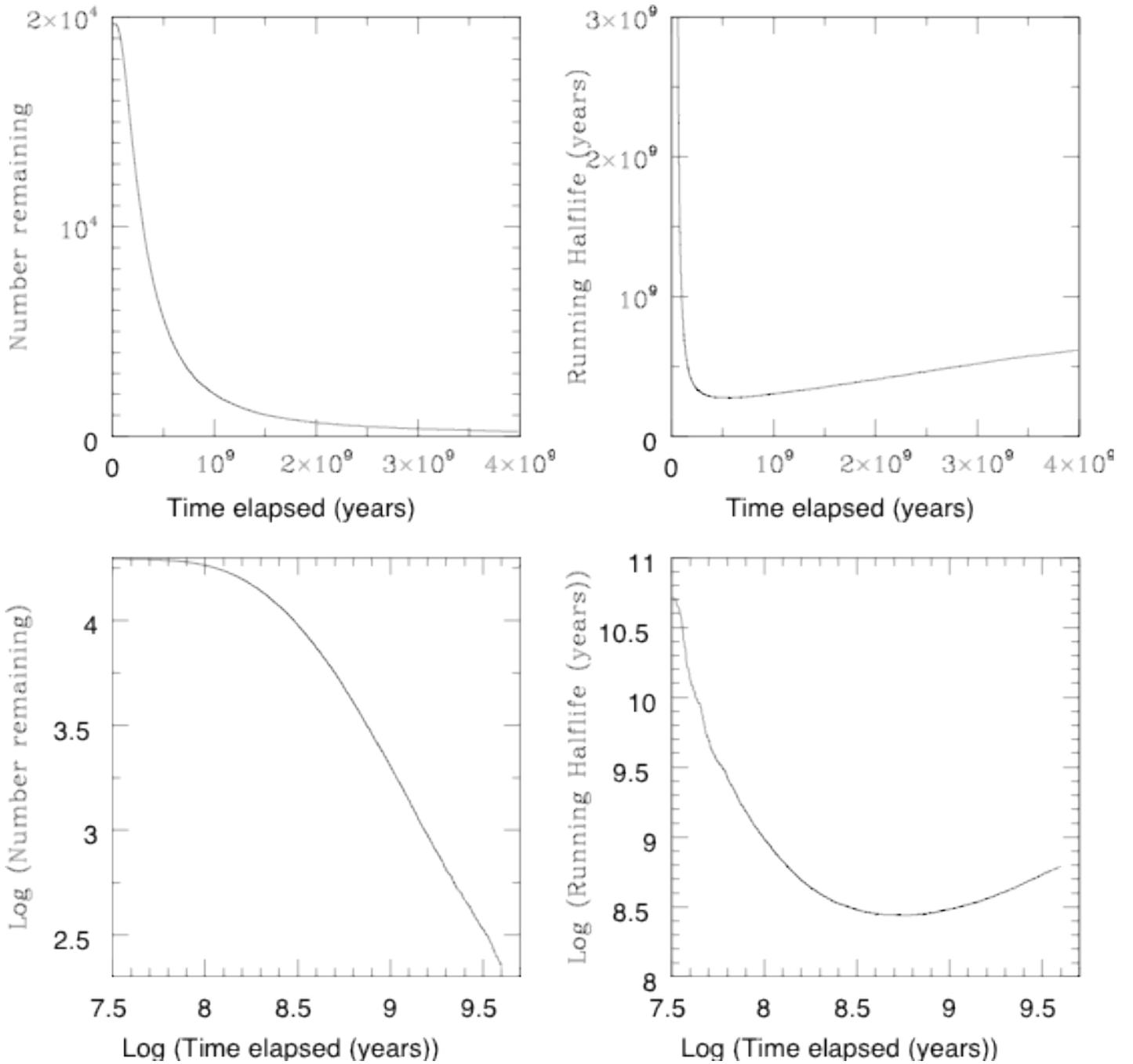

**Figure 3:** The decay of a population of 19683 massless test particles distributed evenly across a region of *a-e-i* element space covering the 3-σ uncertainties in the orbit of (1173) Anchises, as a function of time. The upper-left hand panel shows the decay in the number of surviving test particles as a function of time, while the lower-left hand panel shows the same data in log-log space. The upper-right hand panel shows the illustrative "running half-life" (as described in the text) for the population of test particles, with the lower-right hand panel showing the same information in log-log space.

When the decay of the population of Anchises clones is plotted as a function of time, as can be seen in the upper-left hand panel of Figure 3, it can be well fit by an exponential decay. Following our earlier works, we use this property to calculate a *dynamical half-life* for Anchises of 619 Myr. This value is comparable to those recently obtained for two unstable Neptunian Trojans, 2001 $QR_{322}$ (Horner & Lykawka, 2010b) and 2008 $LC_{18}$ (Horner et al., 2012), and is perfectly compatible with the idea that Anchises is a primordial Jovian Trojan, a surviving member of the unstable component of the Jovian Trojan population which has been gradually decaying since their formation. The remaining panels of Figure 3 highlight different aspects of the decay of Anchises' clones. Although barely visible in the top-left hand panel, it becomes apparent when the decay is plotted in log-log space (lower left) that it takes a certain amount of time for the clones of Anchises to disperse, and for the first test particles to be ejected from the Solar system. This is clearly visible as the plateau at the start of the lower-left hand plot. Similarly, this feature can be seen in the two right hand panels, which show the variation of the "running half-life" for the system as a function of time. The running half-life is a simple illustrative tool that simply shows, at each time, the effective half-life that would be calculated for the suite of test particles based solely on the comparison between the initial population and that remaining at that time. As can be seen, after a short delay, clones of Anchises begin to be ejected from the Solar system in a steady stream, leading to a minimum in the plot of "running half-life" at around the 400 Myr mark, where approximately half the test particles have been removed. From this minimum, at a running half-life of around 350 Myr, the value gradually rises to reach the final calculated value of 619 Myr, at the end of our simulations.

In our previous studies of the dynamically unstable Neptunian Trojans (Horner & Lykawka, 2010b; Horner et al., 2012), we showed that the dynamical stability of those objects was a strong function of their initial orbital elements, with both objects displaying regions of strong dynamical stability and high instability within 3σ of the nominal orbit for the object in *a-e-i* space. It is therefore interesting to look at the distribution of test particles used to examine the behaviour of Anchises to see whether anything similar is happening in this case. Our results are shown in Figure 4 (*a-e* variation in stability) and Figure 5 (*a-i* stability variation).

As can be seen in Figure 4 and 5, the median lifetimes for the test particles considered lie between 200 and 400 million years, regardless of their initial orbital elements. This is not particularly unexpected, however. Unlike 2001 $QR_{322}$ and 2008 $LC_{18}$, which have relatively large uncertainties in their orbital elements, the orbit of Anchises is remarkably well constrained, as a result of ongoing observations spanning more than 80 years. The consistent level of instability across the whole element space considered does, however, reinforce our conclusion that Anchises is truly a dynamically unstable object.

Although the bulk dynamical half-life we calculate for Anchises is 619 Myr, it is probably reasonable to consider that the true dynamical half-life of the object is closer to the value indicated by the minimum of the running half-life plot, namely ~350 Myr, since the final result is somewhat contaminated by the relatively slow decay of the final ~10% of test particles considered. As such, our result suggests a number of possible origins for Anchises.

Firstly, a half-life of ~350 Myr is sufficiently long that it is of course possible that Anchises is a primordial object, one of the last members of a once far greater population of unstable Trojans. If this is indeed the case, and Anchises has been moving on something approximating to its current orbit since the formation of the Trojan population, this would require the initial population of unstable objects to number somewhere between ~5000 to ~20000 times its current value (depending on precisely how lengthy a half-life is assumed, and depending on whether the formation of the Trojan clouds was dated at 4.5 Gyr ago, or during the proposed Late Heavy Bombardment of the Solar system, around 3.8 Gyr ago). Whilst this would require a relatively large population, it is by no means unfeasible.

Second, it is possible that Anchises was originally moving on a significantly more stable orbit, and that that orbit has evolved (and relaxed) over time, with the asteroid gradually random walking into ever less stable regions of the Trojan cloud. Such a mechanism makes perfect sense, since one would expect a gradual flux of material from the more-stable to the less-stable regions of the cloud, as the more stable population gradually decays. Such an evolution could well be aided by collisions between objects in the Trojan cloud (and even collisions between those objects and others on non-Trojan orbits). If Anchises was recently the subject of such a collision, then it is possible that detailed observations of the object could reveal noticeably different properties than those of the bulk of the Trojan population. We note here that no companions have to date been detected in orbit around Anchises, despite its having been surveyed using high angular resolution Adaptive Optics observations (as described by Marchis et al., 2006, the details of which were elucidated by Marchis during the refereeing process). Unfortunately, this means that the bulk density of the asteroid remains unknown. Should future observations lead to the discovery of satellites of Anchises, this might provide evidence in support of such a collisional origin for Anchises' current unstable orbit, as well as providing important additional data on the physical properties of the object.

Finally, the relatively short lifetime of Anchises compared to the age of the Solar system suggests that it might be a relatively recent capture to the Trojan cloud. Since dynamical evolution is a time-reversible process, any region from which objects can escape under purely gravitational evolution can also be reached by such evolution. The more stable the region, the harder it is to escape, and equally, the harder it is for a capture to occur. Since Anchises is relatively loosely bound, compared to the bulk of the Jovian Trojan population, there is always the possibility that it is a relatively recent acquisition to the Trojan cloud (perhaps in the last few hundred million years). Dynamical studies of unstable objects in the outer Solar system have shown that such long-duration captures are certainly possible, even when only a small sample of test objects is considered (e.g. Horner & Evans, 2006). Whilst this is probably the least likely of the three scenarios presented, it is one that future observational work could certainly attempt to address.

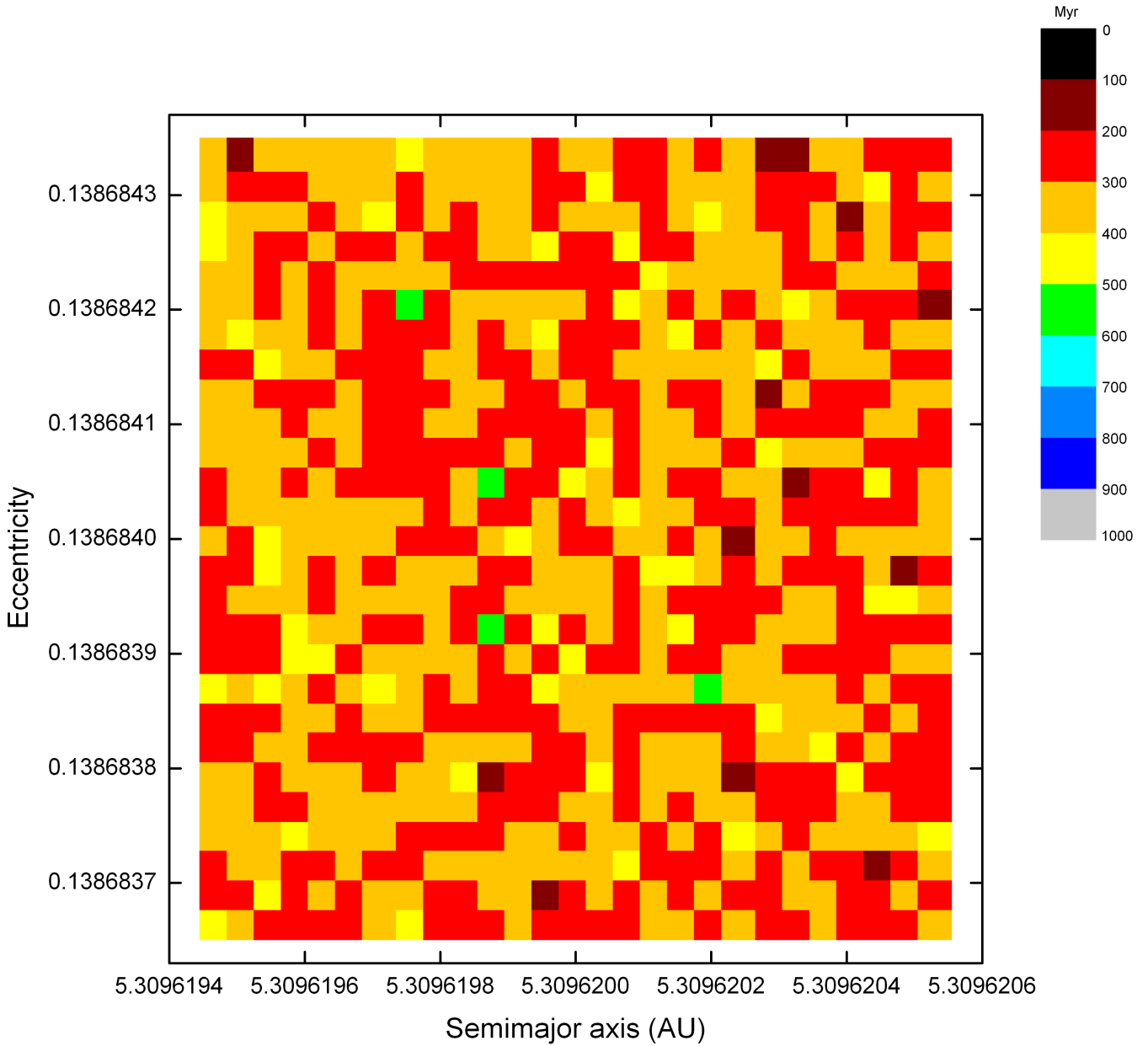

**Figure 4:** The variation in the stability of Anchises as a function of its orbital eccentricity and semi-major axis (AU). Each square in the plot reveals the median lifetime for the 27 test particles that started the simulation with that specific *a-e* combination (the 27 test particles were spread evenly in inclination, spanning ±3σ from the nominal inclination value).

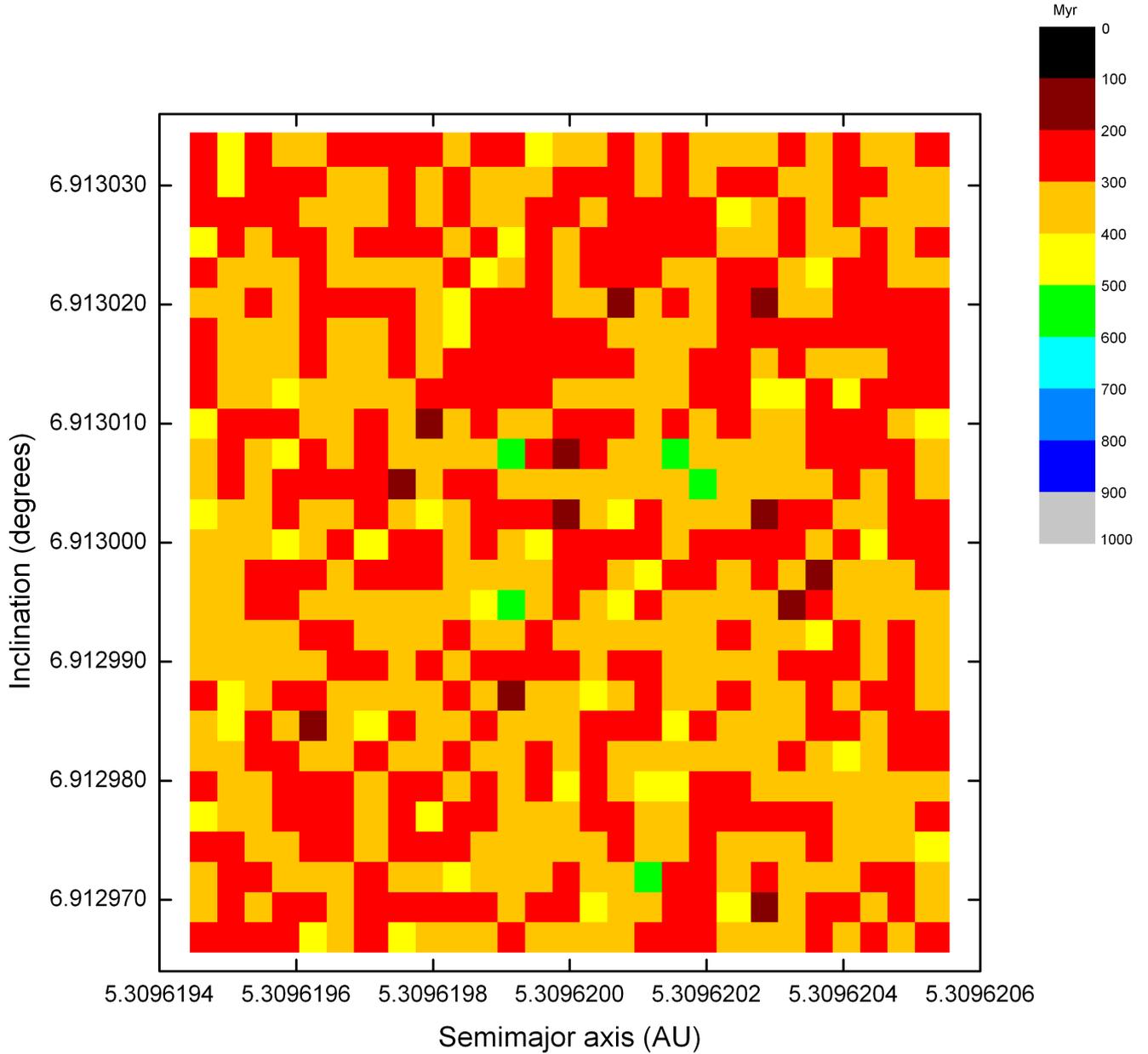

**Figure 5:** The variation in the stability of Anchises' orbit, as a function of its orbital inclination (degrees) and semi-major axis (AU). Each square in the plot reveals the median lifetime for the 27 test particles that started the simulation with that specific *a-i* combination (the 27 test particles were spread evenly in eccentricity, spanning ±3σ from the nominal inclination value).

**Conclusions and Future Work**

We have performed detailed thermophysical and dynamical modelling of the 9$^{th}$ Jovian Trojan to be discovered, (1173) Anchises. Using observations carried out with the space observatories IRAS, Akari, and WISE, together with the optical lightcurve of the object, we have determined that Anchises is an elongated and most likely ellipsoidal body, of dimensions 170 x 121 x 121 km (equivalent in volume to a sphere of diameter $D_{eff}$=137 +18/-11 km). Our modelling reveals that Anchises possesses a relatively high thermal inertia, in the range 25 to 100 $Jm^{-2}s^{-0.5}K^{-1}$ (3-σ confidence interval), one of the largest values measured for any object larger than 100 km in diameter at such large heliocentric distances. This result might be linked to the observed η-values, of around 0.9, and to the small values of the linear phase coefficient describing the absence of strong opposition effects. The albedo determined for Anchises, 0.027 (+0.006/-0.007), makes it one of the darkest objects in the Solar system.

On top of its unexpected physical characteristics, we also find that Anchises exhibits significant dynamical instability on timescales of hundreds of millions of years. Indeed, our simulations showed that fully half of a suite of 19683 Anchises-clones were ejected from the Solar system within just 350 Myr, with only around 1% of the total clone population surviving over the age of the Solar system. Such instability is not unprecedented for planetary Trojans (indeed, the Neptunian Trojans 2001 $QR_{322}$ and 2008 $LC_{18}$ are already known to be dynamically unstable on similar timescales – e.g. Horner & Lykawka, 2010b, Horner et al., 2012), but our results represent a remarkable example of how small body reservoirs can supply objects into unstable orbits in the current Solar system.

Unlike the aforementioned Neptunian Trojans, the dynamical stability of Anchises was not found to vary as a function of the object's initial orbital elements. This is almost certainly the result of the great precision with which Anchises' orbit is known – with an observational arc of 81 years (over 7 full orbits of the Sun), its orbit is far more constrained than either of those objects (which have observational arcs of roughly 5% and 0.5% of one orbit, respectively). As such, the region of orbital element space surveyed for Anchises is far smaller, and so it is certain that the instability observed is a true feature of the object. Such instability does not rule out a primordial origin for Anchises, particularly when one considers that many models of Solar system formation feature planetary migration which would capture objects to the planetary Trojan clouds on a wide variety of orbits (and therefore a wide variety of orbital stabilities). The observed instability could be explained in a number of ways – Anchises could be one of the last members of a once greater dynamically unstable population; it might be a formerly more stable Jovian Trojan that has recently migrated to a less stable region of the Jovian Trojan cloud (potentially as a result of a physical collision with one of its brethren, or simply through chaotic diffusion under the gravitational influence of the Solar system's massive bodies). A less likely scenario, albeit one that is still worth mentioning, is that Anchises could be a relatively recent capture to Jupiter's Trojan population. Such captures have been observed in dynamical simulations of the Centaur population (e.g. Horner & Evans, 2006), albeit on timescales at least two orders of magnitude shorter than the median lifetime we observe for Anchises.

The fact that one of Jupiter's longest known Trojan attendants has been found to be dynamically unstable supports the idea that the planetary Trojans represent a significant source of material to the Centaur population (e.g. Horner & Lykawka, 2010a, c), which are themselves accepted as the proximate parents of the short-period comets (e.g. Horner et al., 2003. 2004a, 2004b). As such, they may well represent a significant contribution to the impact flux at Earth (e.g. Horner & Jones, 2009). For this reason, and given the surprising nature of our thermophysical and dynamical results, Anchises is both a fascinating and beguiling object, and certainly one that merits significant further study.

**Acknowledgements**
JH gratefully acknowledges the financial support of the Australian government through ARC Grant DP0774000. This publication makes use of data products from the Wide-field Infrared Survey Explorer, which is a joint project of the University of California, Los Angeles, and the Jet Propulsion Laboratory/California Institute of Technology, funded by the National Aeronautics and Space Administration. The authors wish to thank the referee of this paper, Franck Marchis, whose comments allowed us to make significant improvements to the paper.